\documentclass[9pt,conference]{IEEEtran}
\IEEEoverridecommandlockouts
\usepackage{amsmath,graphicx,algorithm,algorithmic,adjustbox,setspace}

\usepackage{amsfonts}
\usepackage{textcomp}

\usepackage{xcolor}
\usepackage{enumitem}
\usepackage{hyperref}
\usepackage{multirow}
\usepackage{etoolbox,siunitx}
\robustify\bfseries
\usepackage{booktabs}
\usepackage{balance}
\usepackage{amssymb}
\usepackage{arydshln}
\usepackage{bm}
\usepackage{dsfont}
\usepackage{cite}
\usepackage{tablefootnote}

\usepackage[hang,flushmargin]{footmisc}

\usepackage[inkscapelatex=false]{svg}

\let\OLDthebibliography\thebibliography
\renewcommand\thebibliography[1]{
  \OLDthebibliography{#1}
  \setlength{\parskip}{2pt}
  \setlength{\itemsep}{2pt plus 0.3ex}
}

\def\R{{\mathbb R}}

\title{Leveraging Audio-Only Data\\for Text-Queried Target Sound Extraction
\thanks{This work was performed while K.~Saijo was an intern at MERL.}}

\author{\IEEEauthorblockN{\textit{Kohei Saijo$^{1,2}$, Janek Ebbers$^{1}$, François G.\ Germain$^{1}$,
Sameer Khurana$^{1}$, Gordon Wichern$^{1}$, Jonathan Le Roux$^{1}$}\vspace{.7\baselineskip}}        
\IEEEauthorblockA{{$^{1}$Mitsubishi Electric Research Laboratories (MERL), Cambridge, MA, USA} \;
{$^{2}$Waseda University, Tokyo, Japan}}
}

\begin{document}
\maketitle
\begin{abstract}

The goal of text-queried target sound extraction (TSE) is to extract from a mixture a sound source specified with a natural-language caption.
While it is preferable to have access to large-scale text-audio pairs to address a variety of text prompts, the limited number of available high-quality text-audio pairs hinders the data scaling.
To this end, this work explores how to leverage audio-only data without any captions for the text-queried TSE task to potentially scale up the data amount. A straightforward way to do so is to use a joint audio-text embedding model, such as the contrastive language-audio pre-training (CLAP) model, as a query encoder and train a TSE model using audio embeddings obtained from the ground-truth audio.
The TSE model can then accept text queries at inference time by switching to the text encoder.
While this approach should work if the audio and text embedding spaces in CLAP were well aligned, in practice, the embeddings have domain-specific information that causes the TSE model to overfit to audio queries.
We investigate several methods to avoid overfitting and show that simple embedding-manipulation methods such as dropout can effectively alleviate this issue.
Extensive experiments demonstrate that using audio-only data with embedding dropout is as effective as using text captions during training, and audio-only data can be effectively leveraged to improve text-queried TSE models.

\end{abstract}
\section{Introduction}
\label{sec:intro}

Audio source separation is a fundamental technique for computational scene analysis.
High-fidelity separation has been achieved using neural network (NN)-based approaches~\cite{convtasnet, dprnn, sepformer, saijo2024tf}, pioneered by deep clustering~\cite{dc} and permutation invariant training~\cite{dc,pit}.
While source separation separates all the sources in a mixture, another line of approaches, target sound extraction (TSE), aims to extract only a target source specified by a query, such as a speaker ID~\cite{wang19h_interspeech, vzmolikova2019speakerbeam} or class labels~\cite{ochiai20_interspeech}.
More recently, text-queried TSE~\cite{liu22w_interspeech, kilgour22_interspeech} has been attracting more attention due to its high versatility.

While text-queried TSE has the potential to work on any classes of audios as it accepts natural language as a prompt, realizing this in practice remains a challenge because it would require large-scale high-quality paired text-audio data from many domains. 
As demonstrated in~\cite{liu2023separate}, a text-queried TSE model trained on a small-scale dataset~\cite{liu22w_interspeech} does not generalize to out-of-domain data.
In contrast, AudioSep~\cite{liu2023separate} achieved better generalization than \cite{liu22w_interspeech} by using larger-scale datasets.
Still, results in~\cite{lee2024performance} imply that there is room for improvement by leveraging more data.

Several works have attempted to increase the amount of available data for text-queried TSE.
In~\cite{lee2024performance}, increasing caption variaty using a large language model (LLM)-based caption augmentation has been shown to be effective.
CLIPSep~\cite{clipsep} has been proposed to train a text-queried TSE model using image-audio pairs extracted from videos instead of text-audio pairs, by utilizing a contrastive language-image pre-training (CLIP) model~\cite{radford2021learning} as the query encoder.
The model is trained using images as queries while the text caption is used during inference, which enables the model to be trained without any text captions.
However, using image queries to scale up the data amount may be ultimately ill-suited to TSE due to their intrinsic limitations in capturing out-of-screen and/or background sounds.
Decent performance is achieved at the cost of a complicated training pipeline to estimate such out-of-screen and/or background sounds.

\begin{figure}[t]
\centering
\centerline{\includegraphics[width=\linewidth]{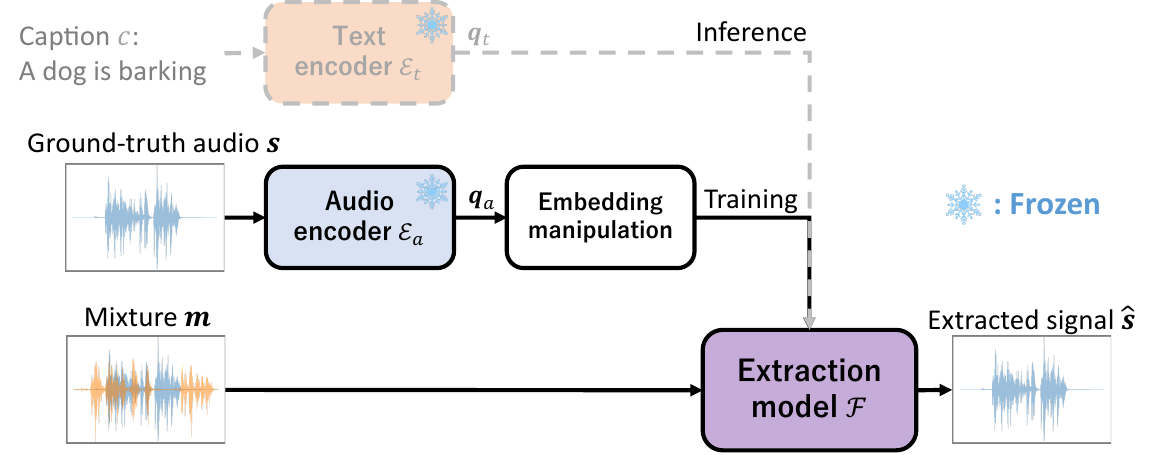}}%
\caption{
    Illustration of text-queried-TSE training with audio-only data.
    Audio embeddings extracted from ground-truth audio by a CLAP audio encoder are used to condition the extraction model during training, while text embeddings are used during inference.
    To prevent the extraction model from overfitting to audio-embedding-specific features, we apply a simple manipulation (e.g., dimension dropout, PCA, etc.) to the embeddings.
}
\label{fig:overview}
\vspace{-4mm}
\end{figure}

To improve the generalizability of text-queried TSE models without being limited by the scarcity and quality of text-audio pairs, a natural approach would be to rely on audio-only data without captions. 
Inspired by CLIPSep, one can think of using a contrastive language-audio pre-training (CLAP)~\cite{CLAP2022} model as the query encoder so that audio embeddings aligned with corresponding text embeddings can be obtained during training.
Although CLAP itself needs a large-scale text-audio pairs, training data can be further scaled up for text-queried TSE if we can leverage audio-only data.
However, including audio-only data in such a way on top of paired text-audio data during training has not been explored so far.
In~\cite{liu2023separate}, an extreme case, where only audio data is used for training, has been investigated but the training failed.
This is likely because of the so-called \textit{modality gap} phenomenon: the two embedding spaces in the CLAP models may not be very well aligned~\cite{liang2022mind}.

Based on the above observation, this work aims to develop ways to effectively leverage audio-only data for text-queried TSE without overfitting to CLAP audio embeddings.
We investigate several methods for manipulating the CLAP audio embeddings, for example, by randomly dropping out dimensions, and show that this can significantly mitigate overfitting.
The experiments demonstrate that audio-queried training with embedding dropout is as effective as normal text-queried training, even for the data that is not used for the CLAP training, which implies that we can potentially scale up the data amount for text-queried TSE.
In addition, we show that such embedding manipulations are effective even in text-queried training and make the TSE model more robust to out-of-domain data.

\section{Audio-queried training for text-queried TSE}
\label{sec:prop}
\subsection{Problem setup}
\label{ssec:problem_setup}

Let us denote the mixture, target source, and non-target source as $\bm{m}$, $\bm{s}$, and $\bm{n} \in \R^{L}$, respectively, where $\bm{m} = \bm{s} + \bm{n}$ and $L$ is the number of samples in the time domain.
The goal of text-queried TSE is to extract $\bm{s}$ using a text caption $c$ that represents the content of $\bm{s}$.
Typically, the caption $c$ is encoded into a $D$-dimensional text embedding $\bm{q}_{t} \in \R^{D}$ with a pre-trained text encoder $\mathcal{E}_{t}$, and $\bm{q}_{t}$ is given to a sound extraction model $\mathcal{F}$ as a conditioning vector:
\begin{align}
  \label{eq:tse}
    \hat{\bm{s}} = \mathcal{F}(\bm{m}, \bm{q}_{t}), \;\;\; \bm{q}_{t}=\mathcal{E}_{t}(c).
\end{align}

To scale up the training data for the text-queried TSE task, we are interested in methods that allow us to effectively incorporate audio-only data during training.
To achieve that goal, we use CLAP~\cite{CLAP2022} to get paired text and audio query encoders $\mathcal{E}_t$ and $\mathcal{E}_a$ (respectively), trained to learn a joint embedding space between text and audio.
As illustrated in Fig.~\ref{fig:overview}, during training, we use audio embeddings $\bm{q}_{a}\in \R^{D}$ obtained by inputting the \textit{ground-truth target source} $\bm{s}$ to the CLAP audio encoder $\mathcal{E}_{a}$, instead of $\bm{q}_{t}$:
\begin{align}
  \label{eq:training}
    \hat{\bm{s}} = \mathcal{F}(\bm{m}, \bm{q}_{a}), \;\;\; \bm{q}_{a}=\mathcal{E}_{a}(\bm{s}).
\end{align}
Such an audio-queried training could in principle work well if the text and audio embedding spaces in the CLAP model were well aligned (i.e., $\bm{q}_t \approx \bm{q}_a $ for pairs of a source $\bm{s}$ and associated caption $c$).
However, in practice, the TSE model overfits to the CLAP audio embedding $\bm{q}_{a}$ because of the modality gap in audio and text sub-spaces, as detailed below. %

\subsection{Modality gap in CLAP}
\label{ssec:modality_gap}

Although CLAP aims to train text and audio encoders to project their inputs to a shared embedding space, it has been shown that the training objective, particularly a low temperature in the contrastive loss, can lead to a \textit{modality gap}~\cite{liang2022mind} between the two embeddings subspaces.
Prior work on CLAP~\cite{CLAP2022, laion-clap, elizalde2024natural} does use a low temperature value (e.g., initialized to 0.007), which suggests that existing CLAP models are likely to suffer from the modality gap.
In addition, intuitively, audio embeddings may have richer information than text ones as text captions do not describe all the content of audios in most cases, which may also cause the gap.
Investigating the average cosine similarity between text and audio embeddings from the 
LAION-CLAP model~\cite{laion-clap} and the Microsoft-CLAP model (MS-CLAP)~\cite{elizalde2024natural} on the AudioCaps dataset~\cite{kim2019audiocaps}, we found that they were indeed only around $0.4$, even though AudioCaps is used for training.

\subsection{Methods for alleviating modality gap}
\label{ssec:methods}

Based on the above discussion, it may be possible to train a CLAP model with a lesser modality gap by using a high temperature. However, higher temperature makes the loss more tolerant to different samples having similar embeddings and thus leads to less discriminative embeddings~\cite{wang2021understanding}, which may cause poor extraction performance in TSE.
We thus focus here on investigating methods to effectively use audio-only data for text-queried TSE without re-training the existing CLAP models.

Although the modality gap exists, relatively low positive cosine similarity (e.g., $0.4$ on AudioCaps) suggests that the embeddings contain both some information shared between the two modalities and some domain-specific information.
To prevent the TSE model from overfitting to such domain-specific features, we explore several methods that make the audio embeddings noisy, by either performing some data augmentation on the audio data or manipulating the audio embeddings.

\textbf{Mixup}: The audio embedding $\bm{q}_a$ is obtained from the mixture of the ground-truth audio $\bm{s}$ with another audio $\bm{s}'$, instead of from $\bm{s}$.
Mixup was shown to be effective in AudioLDM, where a text-queried audio generation model is trained using only audio data~\cite{liu2023audioldm}.
We uniformly sample a signal-to-noise ratio (SNR) between the two audio signals from -5 to +5~dB.

\textbf{SpecAugment}: Some time-frequency (TF) bins of the mel spectrogram are zeroed out.
Frame and frequency masking are done twice, where the number of frames or frequency bins are chosen from [0, 64] or [0, 2], respectively.

\textbf{PCA}: Audio embeddings are projected into a $d$-dimensional space ($d < D$), where the projection matrix is trained with a small amount of text embeddings. We use 1000 text embeddings from the AudioCaps dataset to obtain the projection matrix, and set $d=16$.

\textbf{PCA-inv}: Inverse-PCA is applied after PCA. Unlike normal PCA, the query vector is $D$-dimensional.

\textbf{Gaussian noise}: Scaled Gaussian noise is added to the audio embeddings. It has been shown that this method is effective for alleviating the gap between image and text embeddings in CLIP~\cite{nukrai2022text} or that between text and audio in CLAP~\cite{deshmukh2024training}. Unlike~\cite{nukrai2022text}, we regard the scale $\alpha$ as a hyper-parameter and manipulate embeddings as $\bm{q}_a \xleftarrow{} \bm{q} + \alpha \bm{\upsilon} / ||\bm{\upsilon}||$, where $\bm{\upsilon}$ is zero-mean Gaussian noise. Interestingly, we found that this method improves the robustness of the TSE model even in the text-queried training. Based on preliminary experiments, in each forward pass, we randomly choose $\alpha$ from [1.5, 3.5] when using audio embeddings and [1.0, 2.0] when using text embeddings.

\textbf{Dropout}: $p$-percent of the dimensions of the CLAP embeddings are dropped out (zeroed out) by Bernoulli dropout. The goal is to prevent the extraction model from overfitting to the domain-specific features by removing some information randomly. Based on preliminary experiments, we randomly choose $p$ from [0.75, 0.95] when using audio embeddings and [0.25, 0.75] when using text embeddings.

\section{Related work}
\label{sec:related_work}
This work is inspired by CLIPSep~\cite{clipsep}, where image-audio pairs without captions are leveraged for training text-queried TSE models by utilizing the CLIP encoder.
CLIPSep also suffers from the modality gap in CLIP and fails to train conditional TSE models.
Instead, CLIPSep trains a TSE system which consists of an unconditional separation module and a conditional post-mixing module.
In contrast, our approach is easily applicable to conditional TSE models.
In addition, while complicated training pipeline to estimate out-of-screen sounds is necessary when using an image as a query, we show that audio-queried training with a very simple modification is as effective as text-queried training.

Several prior works on text-queried TSE use CLAP as query encoder.
In~\cite{kilgour22_interspeech} and~\cite{ma2024clapsep}, both text and audio embeddings are used to improve performance or accept audio queries during inference.
While these works assume that they have a text-audio pair for each data, our goal is to leverage audio-only data.
Closest to our work, \cite{liu2023separate} tried audio-only training using the CLAP encoder but the training was not successful\footnote{Note that some results in~\cite{liu2023separate} contradict those in~\cite{kilgour22_interspeech} and ours. The associated code seems to have an issue when extracting audio embeddings, which we suspect is the reason for the contradiction.}.
In contrast, we show that audio-only data can be effectively used with simple embedding manipulation methods.

For the text-queried audio generation task, AudioLDM~\cite{liu2023audioldm} with a CLAP encoder was successfully trained using audio-only data.
The modality gap is alleviated by the mixup augmentation.
In the image field, \cite{nukrai2022text} proposed to train an image captioning model using only text data, where the encoder is a pre-trained CLIP and the decoder is learnable. The modality gap is successfully avoided by injecting Gaussian noise into the CLIP text embedding.
In a similar way, text-only training of an audio captioning model using a pre-trained CLAP has been achieved in~\cite{deshmukh2024training}.
Inspired by these works, we investigate methods to leverage audio-only data in text-queried TSE, which has not been achieved so far.

\begin{table*}[t]
\centering
\sisetup{
detect-weight, %
mode=text, %
tight-spacing=true,
round-mode=places,
round-precision=1,
table-format=2.1,
table-number-alignment=center
}
\caption{
    SI-SDR [dB] of Conformer+LAION-CLAP model on test sets.
}
\vspace{-3mm}
\label{table:main_results}
\resizebox{0.75\linewidth}{!}
{
\begin{tabular}{{cll*{6}{S}S[table-format=2.1,round-precision=1]}}
\toprule

& & & \multicolumn{2}{c}{Captions} & \multicolumn{4}{c}{``This is the sound of \{\textit{class}\}''} \\
\cmidrule(lr){4-5}\cmidrule(lr){6-9}

ID &Method &Training data &{AudioCaps} &{Clotho} &{VGGSound} &{AudioSet} &{ESC50} &{MUSIC} \\

\midrule

\texttt{A0} &- &AC-Text &7.55 &5.10 &6.15 &2.33 &8.33 &0.81 \\
\texttt{A1} &Gaussian noise &AC-Text &7.73 &5.74 &6.83 &2.85 &9.37 &-0.63 \\
\texttt{A2} &Dropout &AC-Text &7.85 &5.77 &6.94 &2.90 &9.21 &0.21 \\

\cmidrule(lr){2-9}

\texttt{A3} &- &AC-Audio &5.05 &1.67 &0.80 &-0.46 &3.62 &-1.20 \\
\texttt{A4} &Mixup &AC-Audio &6.33 &3.15 &4.52 &1.64 &6.55 &2.96 \\
\texttt{A5} &SpecAug &AC-Audio &4.81 &2.97 &-0.28 &-0.50 &3.04 &-0.32 \\
\texttt{A6} &PCA &AC-Audio &6.77 &2.75 &4.68 &0.38 &5.49 &0.05 \\
\texttt{A7} &PCA-inv &AC-Audio &6.93 &3.84 &4.79 &0.63 &6.73 &-0.53 \\
\texttt{A8} &Gaussian noise &AC-Audio &\bfseries 8.05 &6.32 &6.85 &3.42 &9.25 &1.16 \\
\texttt{A9} &Dropout &AC-Audio &\bfseries 8.08 &6.40 &7.03 &3.53 &9.44 &2.02 \\

\cmidrule(lr){2-9}

\texttt{A10} &- &AC-Text-Audio &7.89 &5.95 &6.51 &3.15 &9.05 &1.26 \\
\texttt{A11} &Dropout &AC-Text-Audio &7.97 &6.38 &7.00 &3.53 &9.55 &0.67 \\

\midrule

\texttt{V0} &- &VGG-Text &4.96 &3.97 &8.58 &3.91 &9.49 &9.35 \\
\texttt{V1} &Gaussian noise &VGG-Text &5.83 &5.12 &8.51 &4.45 &9.71 &8.82 \\
\texttt{V2} &Dropout &VGG-Text &6.09 &4.73 &\bfseries 8.76 &4.51 &9.88 &9.22 \\

\cmidrule(lr){2-9}

\texttt{V3} &- &VGG-Audio &4.18 &2.99 &2.34 &1.31 &3.97 &3.50 \\
\texttt{V4} &Mixup &VGG-Audio &6.24 &4.32 &6.05 &2.91 &7.96 &8.31 \\
\texttt{V5} &Gaussian noise &VGG-Audio &7.40 &6.38 &8.42 &4.51 &9.79 &8.61 \\
\texttt{V6} &Dropout &VGG-Audio &7.51 &6.53 &8.54 &4.46 &10.02 &8.66 \\

\cmidrule(lr){2-9}

\texttt{V7} &- &VGG-Text-Audio &7.08 &5.55 &\bfseries 8.78 &4.66 &9.80 &\bfseries 9.59 \\
\texttt{V8} &Dropout &VGG-Text-Audio &7.23 &6.25 &8.53 &\bfseries 5.04 &9.94 &9.04 \\

\midrule

\texttt{AV0} &- &AC-Text + VGG-Text &7.76 &5.86 &8.54 &4.60 &10.42 &8.14 \\
\texttt{AV1} &Dropout &AC-Text + VGG-Text &7.72 &5.97 &8.37 &4.85 &\bfseries 10.21 &8.41 \\

\cmidrule(lr){2-9}

\texttt{AV2} &- &AC-Audio + VGG-Audio &5.44 &2.88 &2.25 &1.38 &4.68 &6.30 \\
\texttt{AV3} &Dropout &AC-Audio + VGG-Audio &7.75 &6.53 &8.40 &4.67 &10.00 &7.94 \\

\cmidrule(lr){2-9}

\texttt{AV4} &- &AC-Text + VGG-Audio &7.89 &5.92 &7.34 &4.22 &9.51 &7.81 \\
\texttt{AV5} &Dropout &AC-Text + VGG-Audio &7.81 &\bfseries 6.56 &8.47 &4.80 &10.07 &8.06 \\

\bottomrule

\end{tabular}
}
\vspace{-4mm}
\end{table*}

\section{Experiments}
\label{sec:experiments}

\subsection{Datasets}
\label{ssec:datasets}
We use the following two datasets for training.
During training, input mixtures are created by uniformly sampling two audio signals and mixing them on the fly, where the signal-to-noise ratio is randomly chosen from -5 to +5 dB.
All the signals are resampled to 32 kHz, following~\cite{liu2023separate}.

\textbf{AudioCaps}~\cite{kim2019audiocaps} includes 10-second audio clips from AudioSet and their human-annotated natural language captions.
AudioCaps was used for the CLAP training~\cite{laion-clap, elizalde2024natural}.
Following the original AudioCaps split, we use 49,827 and 495 audio clips for training and validation, respectively.

\textbf{VGGSound}~\cite{chen2020vggsound} includes pairs of single-class audio clips and their class labels.
Unlike AudioCaps, VGGSound was not used for CLAP training. However, it shares audio clips with AudioSet~\cite{gemmeke2017audio}, which was used to train CLAP, so we removed the shared clips. Thereafter, all mentions of VGGSound refer to the set without the AudioSet clips.
After the filtering, we split the original training data of VGGSound into 169,221 training and 2,500 validation clips.
Since VGGSound does not have natural language captions, we use ``\textit{this is the sound of \{class\}}'' when using text queries.

For testing, we use 6 datasets introduced in~\cite{audiosep_repo}: AudioCaps~\cite{kim2019audiocaps}, Clotho v2~\cite{drossos2020clotho}, VGGSound~\cite{chen2020vggsound}, AudioSet~\cite{gemmeke2017audio}, ESC50~\cite{piczak2015esc}, and MUSIC~\cite{zhao2018sound}.
Each mixture contains two sources from each dataset.
AudioCaps and Clotho v2 have natural language captions, while the other four only have class labels and ``\textit{this is the sound of \{class\}}'' is used as the query.
MUSIC contains instrumental sounds, while the others are mainly composed of environmental sounds.
Please refer to~\cite{liu2023separate} and~\cite{audiosep_repo} for more details.

\subsection{Models}
\label{ssec:models}

As an extraction model, we test two backbones.
Both operate in the short-time Fourier transform (STFT) domain, where a Hann window with length an $L_{w}$~\si{\milli\second} and hop size ${L}_{h}$~\si{\milli\second} is used.

\textbf{Conformer}~\cite{conformer} is the main backbone we use in our investigation as it is reported to work well on environmental sound separation~\cite{selfremixing_scratch} and is light-weight compared with the other model we consider.
The model receives a magnitude spectrogram as input and estimates a real-valued TF mask for the target source.
We use the mixture phase for resynthesis.
It has 16 Conformer encoder layers with 4 attention heads, an attention hidden size of 256, and a feed-forward-network hidden size of 1024.
A conditioning block, composed of a linear layer, Swish activation, a FiLM layer~\cite{perez2018film}, and another linear layer, is placed before each Conformer block to incorporate the conditioning information.
For the STFT, we set $L_w=32$ and $L_h=16$.

\textbf{ResUnet} is a state-of-the-art (SoTA) backbone in text-queried TSE.
We use the same model as in~\cite{liu2023separate}, where FiLM is used as the conditioning layer.
Receiving a complex spectrogram as input, the model estimates a magnitude mask and a phase residual component for the target source, where $L_w=64$ and $L_h=10$.
Please refer to~\cite{liu2023separate} for more details.

As query encoders, we use LAION-CLAP~\cite{laion-clap}\footnote{\url{https://huggingface.co/lukewys/laion_clap/blob/main/music_speech_audioset_epoch_15_esc_89.98.pt}}.
It was pre-trained on a large-scale audio-text dataset and its text encoder is often employed in text-queried TSE~\cite{liu2023separate, ma2024clapsep}.
To examine the effectiveness of audio-queried training on multiple CLAP models, we also test Microsoft-CLAP (MS-CLAP)~\cite{elizalde2024natural}\footnote{\url{https://huggingface.co/microsoft/msclap/blob/main/CLAP_weights_2023.pth}}.

\subsection{Training/evaluation details}
\label{ssec:exp_details}

We train the model for around 400k training steps.
We use the AdamW optimizer~\cite{adamw} with a weight decay factor of 1e-2. %
The learning rate is linearly increased from 0 to 2e-4 in the Conformer and 1e-3 in the ResUnet for the first 4k steps, kept constant for 160k steps, and then decayed by 0.9 every 10k steps.
Gradient clipping is applied with a maximum gradient $L_2$-norm of 5.
The batch size is 32 and the input mixture is 10~\si{\second} long.
As the loss function, we use the negative scale-invariant signal-to-distortion ratio (SI-SDR)~\cite{le2019sdr}.

\subsection{Main results}
\label{ssec:main_results}

Table~\ref{table:main_results} shows the evaluation results of the Conformer+LAION-CLAP model trained on AudioCaps (\texttt{A*}), VGGSound (\texttt{V*}), or both datasets (\texttt{AV*}).
Again, note that AudioCaps has natural language caption and is used for the CLAP training, while VGGSound only has class labels and is not used for the CLAP training.
In \texttt{A10}, \texttt{A11}, \texttt{V7}, and \texttt{V8}, we randomly choose either text or audio query in each training step.

First, compared with the text-queried training (\texttt{A0}), the audio-queried training (\texttt{A3}) gives much worse performance due to the modality gap in the CLAP encoder (see Section~\ref{ssec:modality_gap}).
However, using some data augmentation or embedding manipulation methods can alleviate the problem (\texttt{A4-A9}).
Simple embedding manipulations, namely Gaussian noise and dropout, performed the best among options, and audio-queried training with these manipulations achieves comparable or better performance than text-queried training (\texttt{A0} vs.\ \texttt{A8,A9}).
It is also worth noting that these methods improve the performance of text-queried training on out-of-domain data (\texttt{A0} vs.\ \texttt{A1,A2}).
We believe this is because they augment the text embeddings by adding or removing some noise and have a similar effect to caption augmentation~\cite{lee2024performance}, which makes the TSE model more robust against a variety of captions.
Comparing \texttt{A0} and \texttt{A10}, we observe that using both text and audio as query during training improves the performance over text-only training, which is in line with the results in~\cite{kilgour22_interspeech}.
We again confirm the performance gain when using dropout (\texttt{A10} vs.\ \texttt{A11}), but using both text and audio queries does not improve performance over audio-only training (\texttt{A10} vs.\ \texttt{A9}).

We observe a similar trend when training on the VGGSound dataset: Gaussian noise or dropout on CLAP embeddings is effective in all cases (when using text, when using audio, and when using both).
Comparing \texttt{A0-A2} and \texttt{V0-V2}, we observe that models work best on test data whose caption style (natural language or ``\textit{this is the sound of {class}}'') matches that used in training.
Interestingly, the overfitting to the caption style can be mitigated by using audio queries during training, and overall performance gets better.
This result suggests that using audio as query during training can be beneficial for text-queried TSE even when we have class labels.

Finally, we consider the case where we have a certain amount of text-audio pairs (e.g., AudioCaps) and different audio-only data (e.g., VGGSound).
Results in \texttt{A0} and \texttt{AV4} demonstrate that additional audio-only data helps even without dropout.
Still, dropout contributes to the performance gain (\texttt{AV4} vs.\ \texttt{AV5}) and makes audio-queried training work as well as text-queried training (\texttt{AV5} vs.\ \texttt{AV0}).
This result suggests that we can effectively incorporate audio-only data to improve text-queried TSE systems.

\begin{table}[t]
\centering
\sisetup{
detect-weight, %
mode=text, %
tight-spacing=true,
round-mode=places,
round-precision=1,
table-format=2.1,
table-number-alignment=center
}
\caption{
    SI-SDR [dB] on test sets when training a \textbf{ResUnet}+LAION-CLAP model on AudioCaps.
    \tablefootnote{AudioSep uses the same architecture but trained on a much larger dataset.}
}
\vspace{-2mm}
\label{table:resunet_results}
\resizebox{\linewidth}{!}{
\begin{tabular}{{cc*{6}{S}S[table-format=2.1,round-precision=1]}}

\toprule

& &\multicolumn{2}{c}{Captions} & \multicolumn{4}{c}{``This is the sound of \{\textit{class}\}''} \\
\cmidrule(lr){3-4}\cmidrule(lr){5-8}

Query &Dropout &{AudioCaps} &{Clotho} &{VGGSound} &{AudioSet} &{ESC50} &{MUSIC} \\

\midrule

Text  & &6.27 &3.08 &4.20 &1.53 &5.45 &-0.38 \\
Text  &\checkmark &6.86 &4.18 &6.47 &3.23 &8.41 &0.25 \\
Audio & &2.89 &1.08 &1.11 &0.54 &2.51 &\bfseries 0.78 \\
Audio &\checkmark &\bfseries 7.11 &\bfseries 4.94 &\bfseries 7.24 &\bfseries 3.87 &\bfseries 9.14 &0.55 \\
\midrule
\midrule
\multicolumn{2}{c}{AudioSep~\cite{liu2023separate}} &7.19 &5.24 &9.04 &6.90 &8.81 &9.43 \\

\bottomrule

\end{tabular}
}
\vspace{-3mm}
\end{table}

\subsection{Ablation study}
\label{ssec:results_resunet}

We also trained ResUnet+LAION-CLAP and Conformer+MS-CLAP models to see if we observe a similar trend in other models.
Based on the results of Table~\ref{table:main_results}, we use dropout as embedding manipulation.

Table~\ref{table:resunet_results} shows the evaluation results of ResUnet+LAION-CLAP trained on the AudioCaps data.
The results demonstrate that dropout is also effective on the ResUnet extractor, which implies that audio-queried training with dropout is likely to be effective regardless of the architecture.
We also list the performance of AudioSep~\cite{liu2023separate} since we use the same model architecture\footnote{For fair comparison, all test scores are effectively computed on the same test data as the AudioSep official repository~\cite{audiosep_repo}. However, we do note that the reproduced scores for AudioSep do not match the original paper~\cite{liu2023separate}.}.
Although AudioSep is trained on a much larger dataset, the audio-queried training achieves comparable performance on several test sets, which suggests that we may observe further performance gain over AudioSep by using the same dataset and audio-queried training.

Table~\ref{table:msclap_results} shows the evaluation results of Conformer+MS-CLAP trained on the AudioCaps data.
Again, audio-queried training with dropout gives the best performance.
Although the training data and encoder architectures of MS-CLAP are different from LAION-CLAP, the results show that MS-CLAP also has a modality gap problem and dropout helps to alleviate it.

\begin{table}[t]
\centering
\sisetup{
detect-weight, %
mode=text, %
tight-spacing=true,
round-mode=places,
round-precision=1,
table-format=2.1,
table-number-alignment=center
}
\caption{
    SI-SDR [dB] on test sets when training a Conformer+\textbf{MS-CLAP} model on AudioCaps.
}
\vspace{-2mm}
\label{table:msclap_results}
\resizebox{\linewidth}{!}{
\begin{tabular}{{cc*{6}{S}S[table-format=2.1,round-precision=1]}}

\toprule

& &\multicolumn{2}{c}{Captions} & \multicolumn{4}{c}{``This is the sound of \{\textit{class}\}''} \\
\cmidrule(lr){3-4}\cmidrule(lr){5-8}

Query &Dropout &{AudioCaps} &{Clotho} &{VGGSound} &{AudioSet} &{ESC50} &{MUSIC} \\

\midrule

Text  & &7.49 &5.23 &6.64 &2.18 &8.43 &-0.23 \\
Text  &\checkmark &\bfseries 7.67 &5.69 &7.05 &2.59 &9.01 &1.97 \\
Audio & &3.06 &1.57 &3.76 &0.53 &5.94 &2.14 \\
Audio &\checkmark &7.43 &\bfseries 6.42 &\bfseries 7.81 &\bfseries 3.09 &\bfseries 9.86 &\bfseries 3.33 \\

\bottomrule

\end{tabular}
}
\end{table}

\section{Conclusion}
\label{sec:conclusion}

We investigated methods allowing us to incorporate audio-only data for training text-queried TSE models.
Although the CLAP models often have a modality gap and the TSE models easily overfit to audio queries, simple embedding manipulation methods such as dropout greatly alleviate the problem.
Through experiments using multiple TSE models, we demonstrated that audio-queried training with dropout is as effective as text-queried training.
In future work, we plan to scale up the training data by leveraging large-scale in-the-wild audio-only data.

\clearpage
\balance
\bibliographystyle{IEEEtran}
\bibliography{refs}

\end{document}